\documentclass[preprint2]{aastex631}

\usepackage{hyperref}
\usepackage{amsmath}
\usepackage{multirow}

\graphicspath{{figures/}}

\shorttitle{Type IIP SNe}
\shortauthors{Zha et al.}

\begin{document}

\title{Light Curves of Type IIP Supernovae from Neutrino-driven Explosions of Red Supergiants Obtained by a Semi-analytic Approach}

\correspondingauthor{Shuai Zha}
\email{szha.astrop@gmail.com}

\author[0000-0001-6773-7830]{Shuai Zha}
\affiliation{Tsung-Dao Lee Institute, Shanghai Jiao Tong University, Shanghai 200240, China}

\author[0000-0002-4470-1277]{Bernhard M\"uller}
\affiliation{School of Physics and Astronomy, Monash University, Clayton, Victoria 3800, Australia}
\affiliation{Australian Research Council Centre of Excellence for Gravitational Wave Discovery (OzGrav), Clayton, VIC 3800, Australia}

\author{Amy Weir}
\affiliation{School of Physics and Astronomy, Monash University, Clayton, Victoria 3800, Australia}

\author[0000-0002-3684-1325]{Alexander Heger}
\affiliation{School of Physics and Astronomy, Monash University, Clayton, Victoria 3800, Australia}
\affiliation{Australian Research Council Centre of Excellence for Gravitational Wave Discovery (OzGrav), Clayton, VIC 3800, Australia}
\affiliation{Center of Excellence for Astrophysics in Three Dimensions (ASTRO-3D), Canberra, ACT 2611, Australia}
\affiliation{The Joint Institute for Nuclear Astrophysics, Michigan State University, East Lansing, MI 48824, USA}

\begin{abstract}
Type IIP supernovae (SNe IIP) mark the explosive death of red supergiants (RSGs), evolved massive stars with an extended hydrogen envelope.  They are the most common supernova type and allow for benchmarking of supernova explosion models by statistical comparison to observed population properties rather than comparing individual models and events.  We construct a large synthetic set of SNe IIP light curves (LCs) using the radiation hydrodynamics code \texttt{SNEC} and explosion energies and nickel masses obtained from an efficient semi-analytic model for two different sets of stellar progenitor models.  By direct comparison we demonstrate that the semi-analytic model yields very similar predictions as alternative phenomenological explosion models based on one-dimensional simulations.  We find systematic differences of a factor of $\mathord{\sim}2$ in plateau luminosities between the two progenitor sets due to different stellar radii, which highlights the importance of the RSG envelope structure as a major uncertainty in interpreting LCs of SNe IIP.  A comparison to a volume-limited sample of observed SNe IIP shows decent agreement in plateau luminosity, plateau duration and nickel mass for at least one of the synthetic LC sets.  The models, however, do not produce sufficient events with very small nickel mass $M_\mathrm{Ni}<0.01\,M_\odot$ and predict an anticorrelation between plateau luminosity and plateau duration that is not present in the observed sample, a result that warrants further study.  Our results suggest that a better understanding of RSG stellar structure is no less important for reliably explaining the light curves of SNe IIP than the explosion physics.
\end{abstract}

\section{Introduction} \label{sec:intro}
Core-collapse supernovae (CCSNe) are the spectacular explosions that mark the death of massive stars with zero-age main-sequence masses ($M_{\rm ZAMS}$) greater than $\sim$8--10\,$M_\odot$ \citep[e.g.,][]{Ibeling_13} in the case of single-star progenitors.  Understanding the explosion mechanism of CCSNe has become the equivalent of a millennium problem of modern astrophysics.  CCSNe have great importance as a source of multi-messenger events\footnote{SN 1987A was the first-ever extragalactic astronomical multi-messenger event.} and of compact remnants that they leave behind, and are the origin of most heavy elements in the universe.  

Many of the latest three-dimensional (3D) simulations with sophisticated physics inputs, such as accurate modeling for the neutrino transport, have yielded successful CCSN explosions \citep[see, e.g.,][]{2015ApJ...807L..31L,2017MNRAS.472..491M,2018ApJ...855L...3O,2019MNRAS.485.3153B,2021ApJ...915...28B}, which supports the view that most CCSNe are powered by the neutrino-driven mechanism aided by hydrodynamical instabilities \citep[see the reviews of][]{1990RvMP...62..801B,2012ARNPS..62..407J,2021Natur.589...29B,mueller_20}.  There is, however, still an ongoing discussion on several key issues.  Using phenomenological models, considerable progress has been made in determining how the pre-collapse stellar structure impacts
which stars successfully explode and which ones fail and make black holes \citep{2011ApJ...730...70O,2012ApJ...757...69U,2016ApJ...818..124E,2016MNRAS.460..742M,2016ApJ...821...38S,2015ApJ...801...90P},
but the best structural correlates for ``explodability'' and the parameter space for neutron star and black hole formation are still debated in supernova theory \citep{2020ApJ...890..127C,tsang_20}.
Beyond the question of explodability, both detailed multi-dimensional simulations and phenomenological models have shed
some light on the relation between progenitors and their explosion properties, such as the remnant mass, explosion energy, and nucleosynthesis yields as critical input for chemogalactic evolution \citep{2013ARA&A..51..457N}, but the key challenge is now to more rigorously validate the emerging theoretical picture using observational data.

Direct multi-messenger probes of the CCSN mechanism include gravitational waves (GWs) \citep[e.g.,][]{2011LRR....14....1F,2017hsn..book.1699E,kalogera_19,abdikamalov_22} and neutrinos \citep[e.g.,][]{2017hsn..book.1575J,horiuchi_18,2019ARNPS..69..253M}.  
Current neutrino and GW detectors, however, are only sensitive to events within $\mathord\sim100$\,kpc \citep{2012ARNPS..62...81S,2016PhRvD..94j2001A}.
Electromagnetic signals
are more readily available, especially in today’s
era of large-scale surveys \citep{bellm_14,chambers_16,tonry_18,masci_19,lsst}.

Type~IIP supernovae (SNe IIP) are of particular interest for comparing theoretical CCSN model predictions to observations.
They are the most common observed supernova type and originate from hydrogen-rich red supergiants (RSGs; \citealt{smartt_15}) that are predominantly, though not exclusively, unaffected by binary mass transfer
\citep{podsiadlowski_92,zapartas_21}.  They thus represent the supernova sub-population that most closely matches the progenitor models underlying 
population studies based on phenomenological explosion models\footnote{Note that phenomenological explosions models for stripped stars in binary systems have also been presented recently by \citet{ertl_20} and \citet{schneider_21}.}.
A Type IIP supernova exhibits a $\sim$100-day phase with nearly constant luminosity (``plateau'' -- P) in its light curve (LC) during the inward propagation of a recombination wave through the shock-heated hydrogen envelope.  The plateau luminosity ($L_{\rm pl}$) and duration ($t_{\rm pl}$) are related to the CCSN explosion energy, the progenitor radius, and the mass of the hydrogen envelope \citep{1993ApJ...414..712P,2009ApJ...703.2205K}. The plateau phase is followed by an exponential luminosity tail that is at first powered by the radioactive decays of $^{56}$Ni and $^{56}$Co and by other radioactive species later on.

Historically, supernova explosion and progenitor properties have most often been inferred by fitting \emph{individual} SN LCs with semi-analytic solutions \citep[e.g.,][]{arnett_80,1982ApJ...253..785A} or radiation hydrodynamic simulations \citep[e.g.,][]{2000ApJ...532.1132B,2009ApJ...703.2205K,2011ApJ...729...61B,2010MNRAS.405.2141D}.  This approach, however, can suffer from a degeneracy of the explosion energy and progenitor mass as key parameters that determine the LC \citep{dessart_19}.  The problem of degeneracies can be reduced by considering larger samples of observed transients from surveys or compilations \citep{li_11,faran_14,2015ApJ...799..215P,martinez_22,gutierrez_17,mueller_t_17,martinez_20,martinez_22}.
Most work on inferring progenitor and explosion parameters for larger supernova samples to date has relied on LC fitting, i.e., on reverse modeling \citep{morozova_18,martinez_20}.
The complementary approach is to use forward modeling of entire supernova populations for validating or constraining CCSN explosion models.
Several recent studies produced a considerable number of LCs derived from phenomenological CCSN explosion models \citep{2016ApJ...821...38S,2022ApJ...934...67B,2021ApJ...921..143C}. What is still missing, however, is a global comparison between such a suite of theoretical models and a representative, volume-limited supernova sample. 

Such a comparison also needs to
explore the sensitivity and robustness of explosion parameter and LC predictions to variations in model assumptions. This is particularly important to ascertain the potential for determining physical parameters of individual supernovae or entire populations, e.g., the recent idea
to exploit proposed correlations between iron
core mass and plateau luminosity \citep{2022ApJ...934...67B,2022arXiv221105789B} for use in parameter inference.

In this work, we use the radiation hydrodynamics code \texttt{SNEC} \citep{2015ApJ...814...63M} to calculate LCs of SNe IIP based on two sets of progenitor and explosion models from \citet[][hereafter M16]{2016MNRAS.460..742M} and \citet[][hereafter S16]{2016ApJ...821...38S}.  We obtain explosion properties using the efficient semi-analytic model for neutrino-driven explosions from M16. Though both evolved with the stellar evolution code \texttt{KEPLER} \citep{1978ApJ...225.1021W,2010ApJ...724..341H}, M16 and S16 progenitors have been evolved with slightly different physics assumptions, and illustrate that LC predictions are especially sensitive to model variations that affect the hydrogen envelope.
We then quantitatively compare the models to the volume-limited SNe IIP sample of \citet[][hereafter PP15]{2015ApJ...799..215P} to highlight salient points of agreement and disagreement between the predictions and the observations.  In particular, we highlight that even though the models reproduce the well-known correlation between plateau luminosity $L_{\rm pl}$ and nickel mass $M_{\rm Ni}$, there are still tensions between models and observations in distribution of plateau luminosity, plateau duration and nickel mass.  Similar to \citet{dessart_13}, our results underscore the sensitivity of the LCs to the envelope structure of the progenitor.

Our study has not taken into account the following uncertainties encountered in the theoretical modelling of SNe IIP LCs: the mixing-length parameter and convective overshooting \citep{1987A&A...182..243M,dessart_13}, mixing and composition \citep{2015ApJ...808L..21C,2020A&A...643L..13D}, clumping \citep{2000ApJ...531L.123K,2018A&A...619A..30D,2019A&A...629A..17D}, large scale asymmetry \citep{2015A&A...577A..48W,2021A&A...651A..19D}, line blanketing \citep{2009ApJ...703.2205K,2011MNRAS.410.1739D}, and metallicity \citep{2014MNRAS.440.1856D}. Also our models do not consider the impact of circumstellar materials on SNe IIP LCs \citep{2007ApJ...662.1136C,2019A&A...629A..17D,2022A&A...660L...9D}. How these complexities affect the SNe IIP population warrants further exploration.

This paper is organized as follows.  In \S\ref{sec:model} we review the semi-analytic approach for obtaining neutrino-driven CCSN explosions and the resulting explosion landscape for supernova progenitors.   We compare our approach to two other phenomenological explosion models for the S16 progenitor set in \S\ref{sec:s16}.  In \S\ref{sec:lc} we present the theoretical SN IIP LCs from radiation hydrodynamic simulations, and they are compared to the observational sample in \S\ref{sec:result}.  Our conclusions are given in \S\ref{sec:conclu}.

\section{CCSN explosion model} \label{sec:model}
In this section, we first review the semi-analytic approach of \cite{2016MNRAS.460..742M} for neutrino-driven CCSN explosions.  Then we apply this semi-analytic model to obtain the properties of CCSN explosions for two sets of progenitor models.

\subsection{The semi-analytic approach}

We use the semi-analytic approach of \citet{2016MNRAS.460..742M} to obtain the properties of successful neutrino-driven CCSN explosion such as the explosion energy $E_{\rm exp}$, the baryonic mass of the remnant neutron star $M_{\rm NS,by}$, and the ejected $^{56}$Ni mass $M_{\rm Ni}$. The semi-analytic approach uses physically-motivated scaling laws and solves simple differential equations instead of performing detailed hydrodynamic simulations. A current laptop computer can process nearly $2\mathord,000$ models in just a few minutes. This allows us to explore a large parameter space such as detailed studies in stellar masses. Here, we provide an overview of the treatment of the CCSN dynamics. The full description can be found in \citet{2016MNRAS.460..742M}.

The iron core of a massive star starts to collapse when it reaches a critical mass that depends on its temperature and neutron excess \citep{1968psen.book.....C}.  As the central density reaches nuclear densities, the equation of state stiffens due to nuclear repulsive force,  abruptly halting the collapse. An outgoing bounce shock is launched, but it quickly stalls because of energy losses due to neutrinos and nuclear photo-disintegration. The bounce shock turns into a quasi-stationary accretion shock within a few milliseconds. Material that passes through the shock gets accreted by the proto-neutron star (PNS). Eventually, the shock may be revived to a runaway expansion -- a successful explosion -- or the PNS collapses to a black hole -- a failed explosion. During this accretion phase, copious amounts of neutrinos emanate from the PNS and heat up the matter inside the accretion shock \cite[see, e.g., the reviews in ][]{2012ARNPS..62..407J,2021Natur.589...29B}. The semi-analytic model of \citet{2016MNRAS.460..742M} treats both the \textit{pre-explosion} neutrino heating phase and the subsequent \textit{explosion} phase.

\subsubsection{Pre-explosion phase} 

The region roughly above the PNS and below the accretion shock, dubbed \textsl{gain region}, receives net heating by neutrinos emanating from the PNS due to accretion and PNS cooling.  The gain region is treated as an adiabatically stratified and radiation-dominated layer following \cite{2001A&A...368..527J}. The mass accretion rate $\dot{M}$ is computed following \citet{woosley_15} assuming  that the stellar interior nearly collapses in free fall.  The time evolution of the PNS radius and shock radius and thus the mass in the gain region can be determined from $\dot{M}$ and the mass behind the shock, from which one can, in turn, compute the advection timescale $\tau_{\rm adv}$ and the heating timescale $\tau_{\rm heat}$.  The time of shock revival is determined from the assumption that material must have spent enough time in the gain region for neutrino heating to overcome the binding energy, leading to the critical condition $\tau_{\rm adv}/\tau_{\rm heat}>1$.   If this condition is never met, the model implies that the star forms a black hole.

\subsubsection{Explosion phase} 

During the first episode after shock revival (\textit{Phase I}), outflow and inflow of materials coexist in the post-shock region.  This phase is treated similarly as the \textit{pre-explosion} phase except that the explosion energy $E_{\rm exp}$ is gradually increasing due to the recombination of ejected neutrino-heated material.  The relevant mass outflow rate is computed from the neutrino heating rate and the binding energy at the gain radius based on the heating model from the pre-explosion phase.  As the post-shock velocity (which is computed from the explosion energy, ejecta mass and pre-shock density) exceeds the escape velocity, accretion is assumed to cease, and $E_{\rm exp}$ changes mainly due to explosive nuclear burning and the addition of binding energy of the outer shells (\textit{Phase II}).  We determine $M_{\rm NS,by}$ at the end of \textit{Phase I}, and compute $E_{\rm exp}$ by integration throughout the envelope up to the stellar surface.

The explosive yields of iron-group (IG) elements are computed in a crude way by ``flashing'' shocked material into IG elements when the post-shock temperature exceeds $4.5\times10^9\,$K (instead of $5\times10^9\,$K in the original prescription), but is less than the temperature for $50\%$ dissociation into $\alpha$-particles.  The original model of \citet{2016MNRAS.460..742M} did not account for the contribution of the neutrino-heated ejecta to the IG yields.
To improve upon the original prescription, we take half of these IG elements to be $^{56}$Ni and add another contribution from neutrino-driven outflows, which we assume to be proportional to $E_{\rm exp}$ (as $E_{\rm exp}$ is by construction determined by the amount of ejected neutrino-heated material $M_\nu$), i.e.,
\begin{equation} \label{eq:ni}
    M_\mathrm{Ni} = \frac{1}{2} M_\mathrm{IG}+\frac{1}{2}\alpha E_{\rm exp}
    \approx 
    \frac{1}{2}M_\mathrm{IG}+\frac{1}{2}M_\nu
    \;,  
\end{equation}
where the proportional constant $\alpha$ is set to $m_\mathrm{B}/5\,$MeV.  The second term represents a rough upper limit for the production of nickel by neutrino-driven outflows, corresponding to the optimistic assumption that about half of the neutrino-heated ejecta recombine to $^{56}\mathrm{Ni}$.  We emphasize that an accurate $M_{\rm Ni}$ can only be obtained by multi-D neutrino-transport simulations and that Eq.~(\ref{eq:ni}) only represents a rough estimate. 

Our semi-analytic model includes several parameters that can be used for calibration against more sophisticated multi-D simulations or observational constraints \citep{2015MNRAS.453..287M}, i.e., the shock compression factor, the conversion efficiency of accretion to neutrino luminosity, the PNS cooling timescale (Table~1 of \citealt{2016MNRAS.460..742M}).  These parameters can be used to tune the CCSN explosion landscape, including the explodability and magnitude of $E_{\rm exp}$ considerably.  As a first step, we use the default parameter set and keep the tunability in mind.

Finally, we treat fallback as an all-or-nothing process as in the original prescription \citep{2016MNRAS.460..742M}.  We remark that fallback can significantly influence the properties of explosions for near-critically exploding models. Also, for some failed CCSNe, mass ejection is still possible due to the decrease of the PNS gravitational mass by neutrino emission \citep{2013ApJ...768L..14P,2018MNRAS.476.2366F,2022arXiv220915064S}. However, whereas fallback is now recognized as important for understanding the black-hole mass distribution \citep{mandel_20,mandel_21,antoniadis_22}, these extreme events may not contribute to the SNe IIP population.

\subsection{RSG models and the explosion landscape}
\begin{figure*}
    \centering
    \includegraphics[width=0.95\textwidth]{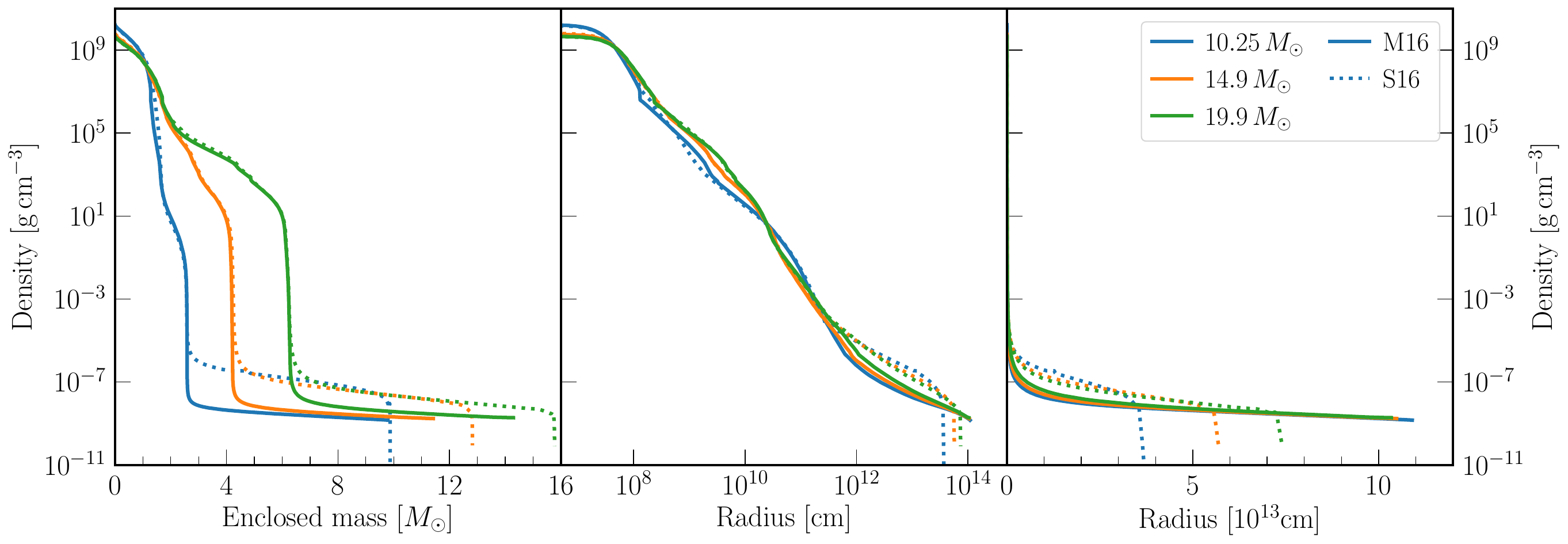}
    \caption{Pre-SN density profiles as a function of enclosed mass (\textsl{left panel}) and radius (\textsl{middle panel}: log scale, \textsl{right panel}: linear scale) for selected progenitor models with $M_\mathrm{ZAMS}=9.5,~14.9$ and $19.9\,M_\odot$ from \citet[][M16, \textsl{solid lines}]{2016MNRAS.460..742M} and \citet[][S16, \textsl{dotted lines}]{2016ApJ...821...38S}. All the models successfully explode. In particular, 9.5\,$M_\odot$ is the minimum mass common to both sets, and 14.9$M_\odot$  and 19.9\,$M_\odot$ are the
    closest progenitor masses to 15\,$M_\odot$ and 20\,$M_\odot$ with explosions in both sets. }
    \label{fig:prog_eg}
\end{figure*}

\begin{deluxetable*}{cccccccccccc}
    \tablecaption{Presupernova, explosion and light-curve properties for the progenitor models shown in Fig.~\ref{fig:prog_eg}.}   
    \label{tab:prog_eg}
    \tablehead{\colhead{$M_{\rm ZAMS}$} & \colhead{Source} & \colhead{$M_{\rm prog}$} & \colhead{$R_{\rm prog}$} & \colhead{$M_{\rm Fe}$} & \colhead{$M_{\rm env}$} & \colhead{$\xi_{2.5}$} &  \colhead{$E_{\rm exp}$} & \colhead{$M_\mathrm{Ni}$} & \colhead{$M_\mathrm{NS,by}$} & \colhead{$L_\mathrm{pl}$} & \colhead{$t_\mathrm{pl}$}
    \\
    \colhead{($M_\odot$)}  & {} & \colhead{($M_\odot$)} & \colhead{($10^{13}$\,cm)}& \colhead{($M_\odot$)} & \colhead{($M_\odot$)} & {} & \colhead{($10^{51}$\,erg)} & \colhead{($10^{-2}\,M_\odot$)} & \colhead{($M_\odot$)} & \colhead{($10^{8}\,L_\odot$)} & (days)
    }
    \startdata 
    \multirow{2}{*}{9.5} 
     & M16 & 9.11 & 10.19 & 1.29 & 6.77 & $1.6\times10^{-5}$ & 0.25 & 2.3 & 1.35 &   3.37 & 144 \\
     & S16 & 9.16 & 2.87 & 1.30 & 7.13 & $6.1\times10^{-5}$ & 0.32 & 2.8 & 1.34 & 1.48 & 114 \\ \hline
    \multirow{2}{*}{14.9}  
     & M16  & 11.4 & 10.5 & 1.56 & 7.27 & 0.15 & 0.99 & 5.2 & 2.18 & 10.9 & 96 \\
     & S16  & 12.8 & 5.70 & 1.50 & 8.62 & 0.16 & 1.06 & 5.5 & 2.18 & 6.19 & 95\\ \hline
    \multirow{2}{*}{19.9} 
     & M16  & 14.3 & 10.3 & 1.56 & 8.18 & 0.20 & 1.20 & 11.0 & 1.74 & 11.3 & 95 \\
     & S16  & 15.8 & 7.41 & 1.53 & 9.63 & 0.22 & 1.32 & 11.4 & 1.76 & 8.42 & 96 \\ \hline
    \enddata
    \tablecomments{Here, $M_{\rm ZAMS}$ is the ZAMS mass for the pre-SN model. M16 and S16 stand for progenitor from the sets of \cite{2016MNRAS.460..742M} and \cite{2016ApJ...821...38S}, respectively.  $M_{\rm prog}$ and $R_{\rm prog}$ are the stellar mass and radius, $M_{\rm Fe}$ and
    $M_{\rm env}$ are the masses of the iron core and hydrogen envelope, and  $\xi_{2.5}$ is the compactness (Eq.~\ref{eq:xi}), all defined at the onset of collapse. $E_{\rm exp}$, $M_{\rm Ni}$ and $M_{\rm NS,by}$ are the resulting explosion energy, nickel mass and remnant neutron-star mass obtained by the semi-analytic model of \cite{2016MNRAS.460..742M}.  $L_{\rm pl}$ and $t_{\rm pl}$ are the plateau luminosity and duration of the resultant SN IIP light curve obtained by \texttt{SNEC} simulations. }
\end{deluxetable*}

We apply the semi-analytic approach to two sets of single-star solar-metallicity RSG models as CCSN progenitors, which we refer to as M16 \citep{2016MNRAS.460..742M} and S16 \citep{2016ApJ...821...38S}.  Both sets were evolved with the stellar evolution code \texttt{KEPLER} \citep{1978ApJ...225.1021W,2010ApJ...724..341H} but with two major known differences in the physical inputs.  One is that the erroneous pair-neutrino loss rate was updated to a corrected version in M16 but not in S16 \citep[see \S2 of ][]{2018ApJ...860...93S}.  This can affect the late burning stages after core helium depletion.  The other difference is that a fixed, large boundary pressure was used at the stellar surface in M16 to keep the models stable.  This affected the RSG structure, making them more compact and affecting the mass loss during the reg giant phase.  Other differences may exist, such as the helium burning rates that impact the size of the carbon oxygen core after core helium depletion\citep{Imbriani_01,Tur_07,West_13}.

\begin{figure}
    \centering
    \includegraphics[width=0.47\textwidth]{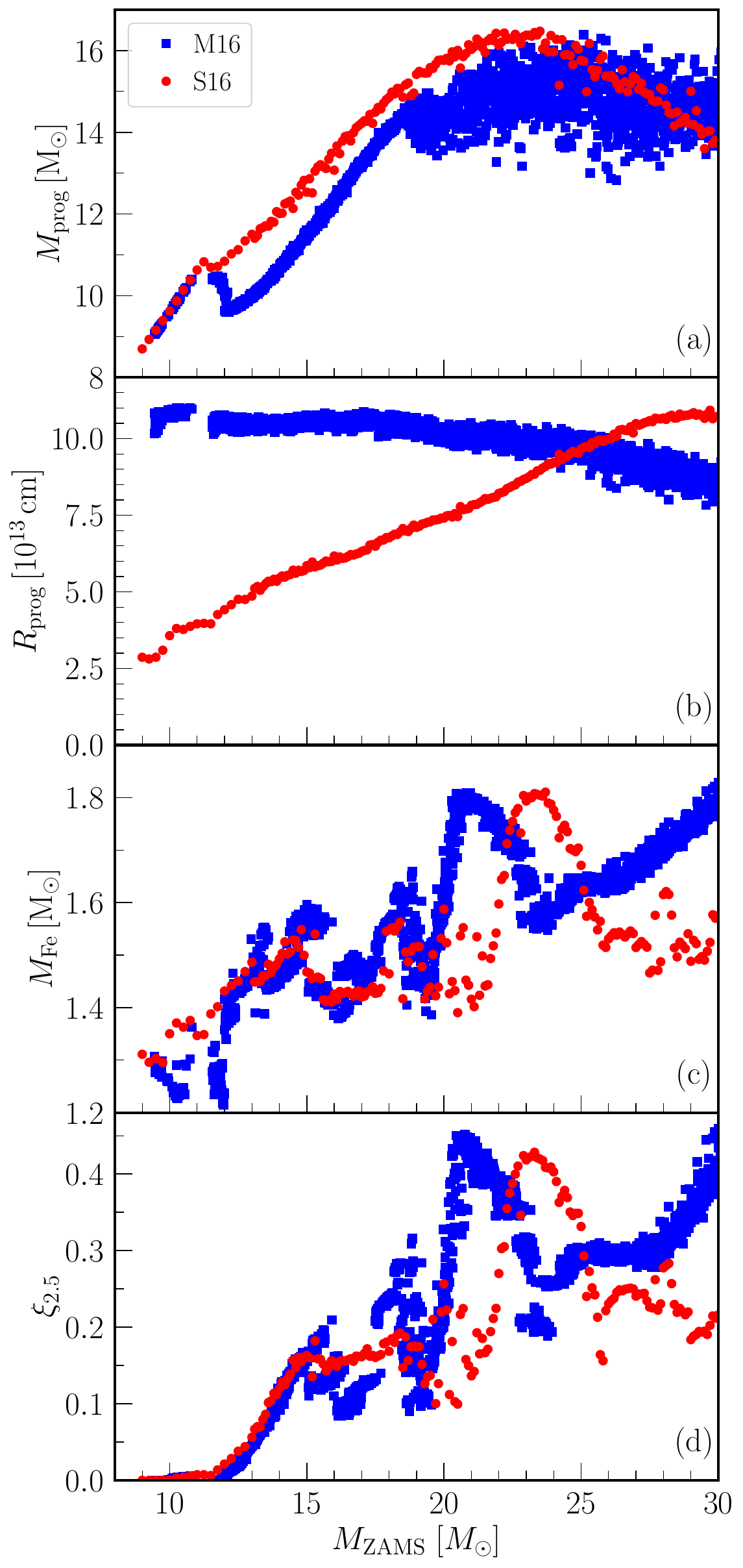}
    \caption{Comparison of progenitor properties as a function of ZAMS mass between M16 (\textsl{blue squares}) and S16 (\textsl{red dots}) models. From top to bottom, the panels show the pre-SN stellar mass ($M_\mathrm{prog}$), pre-SN stellar radius ($R_{\rm prog}$), iron-core mass ($M_\mathrm{Fe}$) and compactness parameter ($\xi_{2.5}$) at the onset of collapse. A choice of $\xi_{2.5,\rm crit}=0.263$  $(0.243)$ best discriminates the explodability for the M16 (S16) models with 150 (28) false identifications.}
    \label{fig:prog_props}
\end{figure}

The differences between the two sets at the onset of collapse are shown by the comparison of the pre-SN density profiles for three selected values of $M_{\rm ZAMS}$ (Fig.~\ref{fig:prog_eg}), and the global parameters for the pre-SN stellar structure (Fig.~\ref{fig:prog_props}).  Figure~\ref{fig:prog_eg} clearly shows that the larger pressure cut results in a more dilute hydrogen envelope for M16 models whereas the core structures are nearly the same. This can also be inferred from the larger pre-SN stellar radii $R_\mathrm{prog}$ for M16 models with $M_{\rm ZAMS}\lesssim24\,M_\odot$ (panel~(b) of Fig.~\ref{fig:prog_props})\footnote{The M16 models radii are inflated due a finite-pressure boundary condition that was set to ensure stellar stability. Note that, as is often the case in stellar structure, the response to a finite surface pressure can be non-intuitive, in this case resulting in expansion instead of contraction.}, although the different pre-SN stellar masses ($M_\mathrm{prog}$, panel~(a) of Fig.~\ref{fig:prog_props}) also indicate subtle differences in the mass loss rates as a result of feedback processes that requires further study but is beyond the scope of this work.  A striking difference is the opposite trends of $R_\mathrm{prog}$ versus $M_{\rm ZAMS}$.  The progenitor radius, $R_\mathrm{prog}$ is positively correlated with progenitor mass in the S16 models, but decreases slightly with mass in the M16 models.

Figure~\ref{fig:prog_props} also illustrates differences in the core structure between the two sets.  The S16 models have a smaller mass of the carbon-oxygen core than M16 models for the same $M_{\rm ZAMS}$, which carries through to later evolutionary phases.  This is reflected by the final iron-core mass $M_\mathrm{Fe}$ (panel~(c) of Fig.~\ref{fig:prog_props}), and can also be inferred from the progenitor compactness $\xi_{2.5}$ (panel~(d) of Fig.~\ref{fig:prog_props}).  Here $\xi_{2.5}$ is defined as \citep{2011ApJ...730...70O}
\begin{equation} \label{eq:xi}
    \xi_{M} = \frac{M/M_\odot}{R(M_{\rm baryon}=M)/1\mathord,000\,\mathrm{km}} \bigg|_{t=t_0}\;,
\end{equation}
where $M$ is set to be $2.5\,M_\odot$, and $t_0$ is the time at the onset of collapse defined as when the infall speed anywhere in the core first exceeds $10^{8}$\,cm\,s$^{-1}$.  Structures in the landscape of $\xi_{2.5}$ are systematically shifted to higher $M_\mathrm{ZAMS}$ in the S16 models.  Except for this shift, M16 and S16 models have quite similar core structures, with stochastic variations in $\xi_{2.5}$ for $M_\mathrm{ZAMS}\simeq15$--$20\,M_\odot$ due to the chaotic merging of oxygen and carbon- and neon-burning shells \citep{2018ApJ...860...93S,2018MNRAS.473.1695C,2020ApJ...890...94Y}. The impact of the erroneous neutrino loss rate is most significant for stars with $M_{\rm ZAMS}\gtrsim20\,M_\odot$, which constitute only $\sim18\%$ of all the progenitors and even less for exploding models.  Therefore, the overall impact on the ensemble of SNe IIP LCs is small.

\begin{figure}[t!]
    \centering
    \includegraphics[width=0.47\textwidth]{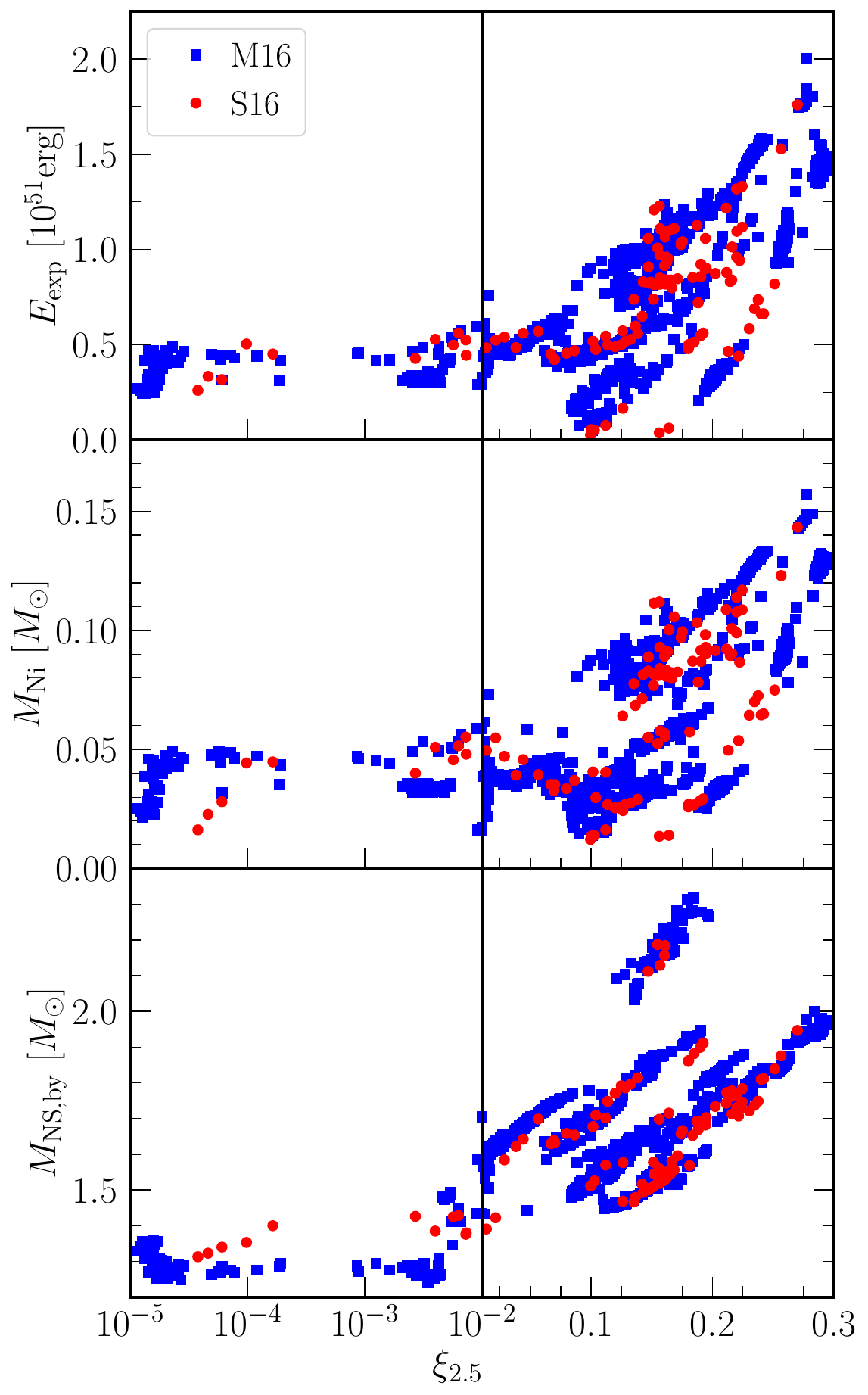}
    \caption{Similar to Fig.~\ref{fig:prog_props}, but for the  comparison of explosion properties as a function of progenitor compactness $\xi_{2.5}$ between the M16 (\textsl{blue squares}) and S16 (\textsl{red dots}) models. From top to bottom, the panels show the explosion energy $E_{\rm exp}$, $^{56}$Ni mass $M_{\rm Ni}$ and baryonic neutron star mass $M_{\rm NS,by}$. Note that for $\xi_{2.5}\le0.01$ we use a log scale and for $\xi_{2.5}>0.01$ we use a linear scale.}
    \label{fig:exp}
\end{figure}

We only consider pre-SN models with $M_{\rm ZAMS}\le 30\,M_\odot$, because models with a larger $M_{\rm ZAMS}$ would exceed the Humphreys-Davidson limit and experience significant mass loss and result in SNe other than type IIP,  aside from the fact that few explosions are predicted in this region in the first place.  For M16 we have 1891 models with a mass resolution of 0.01\,$M_\odot$, for which 991 successfully explode.  For S16 we have 187 models with a mass resolution of 0.1\,(0.25)\,$M_\odot$ at $M_\mathrm{ZAMS}$ above (below) 13\,$M_\odot$, for which 115 models successfully explode.  In Fig.~\ref{fig:exp} we show the explosion properties predicted by the semi-analytic supernova model
as a function of the $\xi_{2.5}$.  We find good agreement between the two sets of progenitors and determine a critical $\xi_{2.5} = 0.263$ ($0.243$) that best discriminates the explodability for M16 (S16) models. 

\section{Comparison of alternative phenomenological explosion models (S16 set)} 
\label{sec:s16}

\begin{figure}
    \centering
    \includegraphics[width=0.47\textwidth]{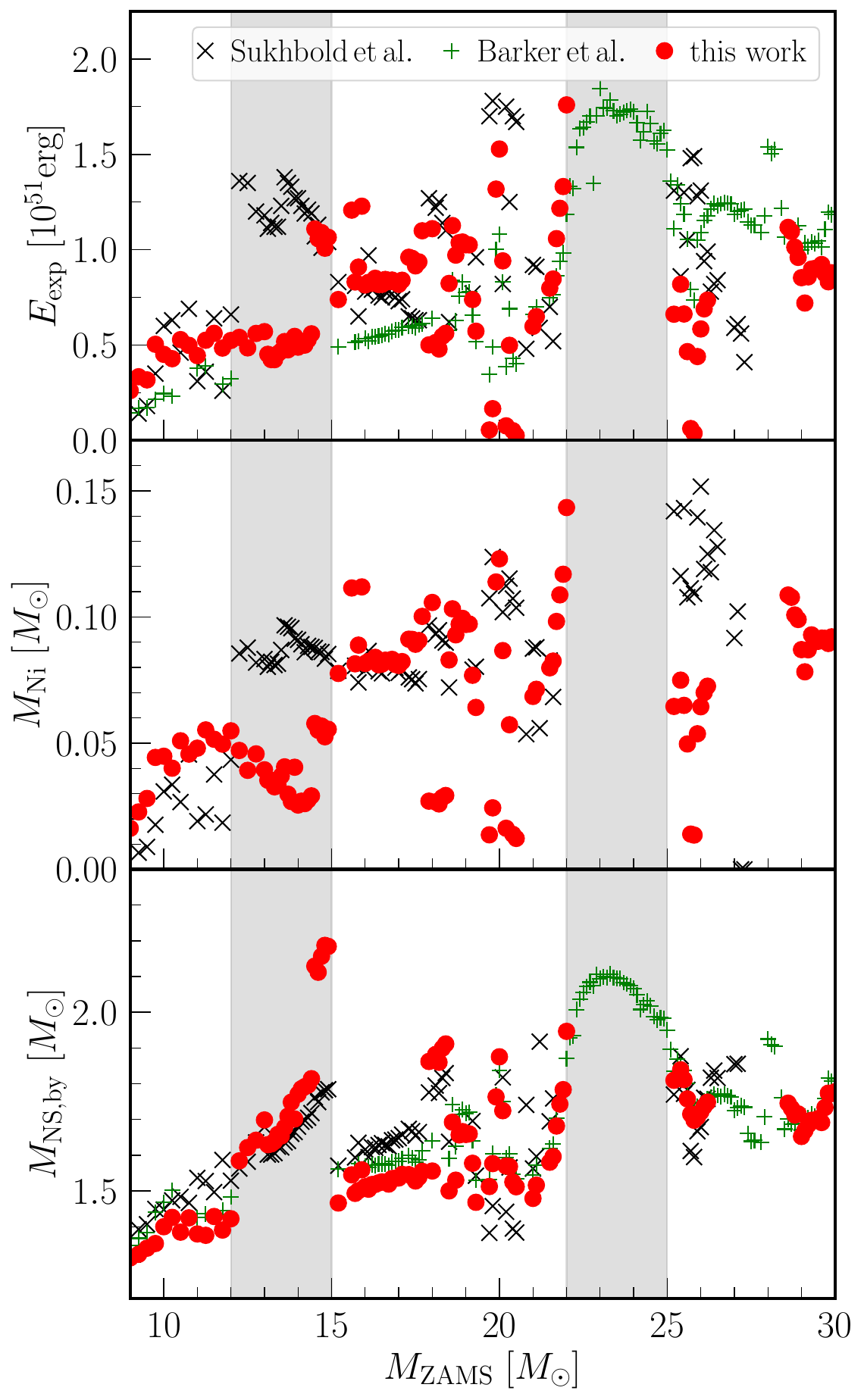}
    \caption{Comparison of the explosion properties of S16 progenitors as obtained in \citeauthor{2016ApJ...821...38S} (\textsl{black crosses}), \citeauthor{2022ApJ...934...67B} (\textsl{green open squares}) and this work (\textsl{red dots}).  Note that \citeauthor{2022ApJ...934...67B} did not calculate $M_\mathrm{Ni}$ but used the nickel masses of S16 instead. }
    \label{fig:exp_S16}
\end{figure}

It is currently not feasible to perform 3D simulations with neutrino transport to determine the properties of CCSN explosions for a sufficiently large number of progenitors required for population studies.  Our semi-analytic model is among several efficient phenomenological approaches to predict the outcome of collapse (explosion or non-explosion) as well as explosion and remnant properties \citep{2011ApJ...730...70O,2012ApJ...757...69U,2015ApJ...801...90P,2015ApJ...806..275P,2016ApJ...821...38S,2020ApJ...890..127C,ertl_20,2022ApJ...934...67B,ghosh_22}.  Most other studies rely on 1D simulations that mimic the supportive role of multi-dimensional flow instabilities in enabling shock revival either by increasing the neutrino emission, the neutrino energy deposition, or by means of 1D turbulence models (but see \citealt{mueller_19} for a critical discussion of this approach).  Qualitative and quantitative differences and similarities between the various phenomenological models have been discussed in the literature, and \citet{pejcha_20} also provides a side-by-side comparison of important outcomes such as the relation between explosion energy and nickel mass or the predicted neutron star mass distribution.  Such comparisons can be somewhat skewed by differences in the size, mass range, and input physics of underlying stellar evolution model sets.
 
For this reason, it is useful to compare our results to those obtained by different 1D simulation studies for the S16 progenitor set, namely from the study of \citeauthor{2016ApJ...821...38S} \citep{2016ApJ...821...38S} and \citeauthor{2022ApJ...934...67B} \citep{2022ApJ...934...67B}.  \citeauthor{2016ApJ...821...38S} used the \texttt{P-HOTB} code \citep{2012ApJ...757...69U,2016ApJ...818..124E} with a gray neutrino-transport scheme and a proto-neutron star core model, and is calibrated by two well-observed CCSNe.  Their models are calibrated to inferred explosion properties for SN 1054 and SN 1987A at the respective progenitor masses. The SN~1987A calibration is used for all  progenitors with $M_{\rm ZAMS}>12\, M_\odot$, and for $M_{\rm ZAMS}<12\, M_\odot$ interpolation between the relevant model parameters for the two calibration cases is applied. \citeauthor{2022ApJ...934...67B} used the \texttt{FLASH} code with a multi-group two-moment neutrino-transport scheme \citep{2018ApJ...865...81O} plus the \texttt{STIR} method for simulating turbulence in 1D \citep{2020ApJ...890..127C}.  Their \texttt{STIR} method is calibrated to fit full 3D simulations run in the same code \citep{2018ApJ...865...81O}.

The comparison is shown in Fig.~\ref{fig:exp_S16} for $E_{\rm exp}$, $M_{\rm Ni}$ and $M_{\rm NS,by}$.  Although with quite different implementations and degrees of approximations, we find considerable agreements among the results from \citeauthor{2016ApJ...821...38S}, \citeauthor{2022ApJ...934...67B} and this work. 
The agreement is especially remarkable for the baryonic neutron star mass $M_{\rm NS,by}$, which once again confirms the important role of
the Si-O shell interface as a natural point
for the onset of the explosion and a strong predictor for the final neutron star mass.

Discrepancies are noteworthy mainly in the mass ranges with near-critical explodability (gray shaded bands in Fig.~\ref{fig:exp_S16}), with $M_{\rm ZAMS}\simeq$ $12$--15\,$M_\odot$ and $22$--25\,$M_\odot$.  For $M_{\rm ZAMS}\simeq$ $12$--15\,$M_\odot$, \citeauthor{2022ApJ...934...67B} predicts no explosion while both \citeauthor{2016ApJ...821...38S}
and the semi-analytic model obtain explosions.
$E_{\rm exp}$ and $M_{\rm Ni}$ in \citeauthor{2016ApJ...821...38S} are, however, larger by about a factor of 2.5 than those in this work, which may be related to the change in calibration case of \texttt{P-HOTB} from SN 1054 to SN 1987A at $12\,M_\odot$. 

On the other hand, for $M_{\rm ZAMS}\simeq$ $22$--25\,$M_\odot$, \citeauthor{2016ApJ...821...38S} and our semi-analytic model predict no explosion, whereas \citeauthor{2022ApJ...934...67B} yields relatively large explosion energies $E_{\rm exp}$ ($\ge2\times10^{51}$\,erg).  The explodability of these critical models is still under debate with state-of-the-art 3D simulations \citep[e.g.,][]{2018ApJ...855L...3O,2020ApJ...891...27M,2020MNRAS.491.2715B}. The mass distribution of observed SN IIP progenitors \citep{smartt_15} and first observational evidence for the quiet disappearance of a RSG \citep{adams_17}, presumably by stellar collapse favor a lower probability of explosion in this mass range\footnote{A potential exception is SN 2015bs whose progenitor was inferred to have a metallicity of $\le0.1Z_\odot$ and a $M_{\rm ZAMS}$ of 17-25\,$M_\odot$ \citep{2018NatAs...2..574A}}..

The overall trends and patterns in explosion energy are qualitatively compatible between the three phenomenological models outside the gray-shaded areas. They all predict low explosion energies at the low-mass end, a general trend towards higher explosion energies in the range of $15$--22\,$M_\odot$ with considerable scatter at higher masses.  Above 25\,$M_\odot$, the agreement is less convincing. It is noteworthy, however, that even in the region of $22$--25\,$M_\odot$, where \citeauthor{2022ApJ...934...67B} disagrees qualitatively with the other two models, the high explosion energies reflect
a similar pattern in \citet{2016MNRAS.460..742M} with parameter choices that increase explodability (i.e., higher turbulent pressure in the gain region or a higher accretion efficiency for neutrino emission).

The situation for the nickel masses, $M_\mathrm{Ni}$, which are only available for \citeauthor{2016ApJ...821...38S} and our semi-analytic model, is similar to the explosion energies.  There is rather good agreement between \citeauthor{2016ApJ...821...38S} and our work below $22\,M_\odot$, which is rather striking considering the relatively simple model for nickel production used in our approach.

These results demonstrate that predictions of explosion and remnant properties from the three phenomenological models are quite robust to differences in the methodology, once some form of calibration (e.g., for one or two specific supernovae or for the typical energy range of observed explosions) is applied.

\section{Theoretical light curves of Type IIP SNe} \label{sec:lc}

With the explosion properties ($E_{\rm exp}$, $M_{\rm Ni}$ and $M_{\rm NS,by}$) obtained in \S\ref{sec:model}, we utilize \texttt{SNEC} \citep{2015ApJ...814...63M} to generate LCs of SNe IIP from M16 and S16 progenitors. \texttt{SNEC} is an open-source spherically-symmetrical radiation hydrodynamics code with the capability to follow the shock propagation through the stellar envelope.  It solves the Lagrangian hydrodynamics equations supplemented with a radiation diffusion term.  Note that \texttt{SNEC} assumes local thermal equilibrium between matter and radiation, which fails during the shock breakout
and nebular phase, but is reasonably reliable for LCs during the plateau phase \citep{1993A&A...273..106B} that is of interest here.  We refer to the code paper \citep{2015ApJ...814...63M} and documentation \citep{SNEC} for details on the numerical implementation.

We employ the default settings of \texttt{SNEC}, such as the equation of state, ionization treatment and opacities.  The newborn NS with $M_{\rm NS,by}$ is excised from the numerical grid and a thermal bomb is used to initialize the shock.  The sum of $E_{\rm exp}$ and binding energy of the mass content above the excised NS is spread into the $0.1\,M_\odot$ above the excised boundary so that the final explosion energy equals the desired value $E_{\rm exp}$ \citep{SNEC}.  For the mixing of nickel, we simply spread $M_{\rm Ni}$ homogeneously up to $3\,M_\odot$ as our semi-analytic approach cannot treat the mixing.  The mixing of nickel is beyond the scope of this paper but its impact on SNe IIP LCs may be worth further investigation (see, e.g., \citealp{utrobin_17}).  We evolve all the models to $\sim200$ days, by which time all models have reached the radioactively-powered tail phase.  For comparison to observations, we are particularly interested in two LC parameters: the plateau luminosity $L_{\rm pl}$ and the plateau duration $t_{\rm pl}$.  We take the bolometric luminosity at 50 days after the shock break out as $L_{\rm pl}$. The determination of $t_{\rm pl}$ is more tricky; we tentatively pick the time of the steepest gradient of the $B$-band magnitude as the end of plateau phase.  We also present the photospheric velocity at 50 days $v_{50}$ after the shock breakout, which is a proxy for the mean expansion rate of the ejecta and $E_{\rm exp}/M_{\rm ejecta}$, with $M_{\rm ejecta}$ being the total mass of the ejecta. Here we use the \texttt{SNEC} definition for the location of photosphere, i.e. by the optical depth $\tau=2/3$. The key LC and explosion parameters for all models are publicly available at Zenodo: doi:\href{https://doi.org/10.5281/zenodo.7354733}{10.5281/zenodo.7354733} in the same form as listed in Table~\ref{tab:prog_eg} .

\begin{figure}
    \centering
    \includegraphics[width=0.48\textwidth]{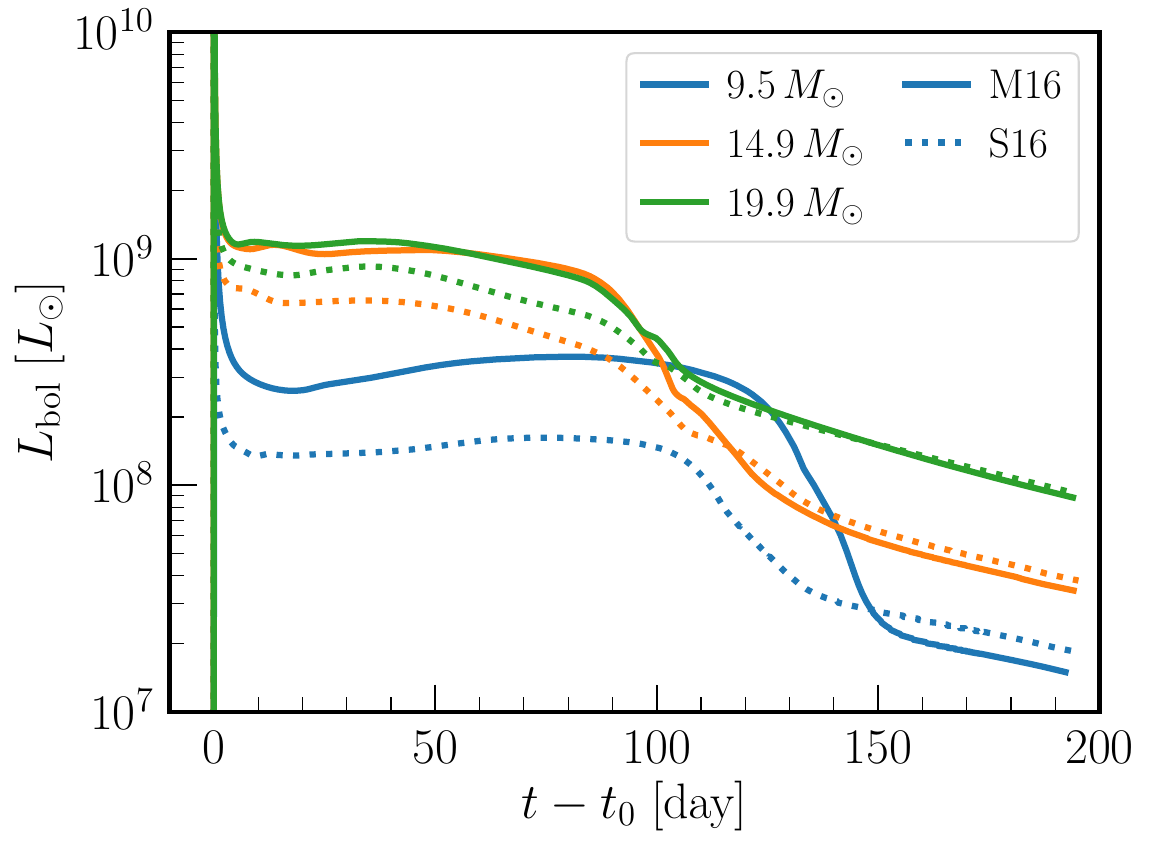}
        \caption{Bolometric light curves of SNe IIP from the M16 (\textsl{solid lines}) and S16 (\textsl{dotted lines}) pre-SN models shown in Fig.~\ref{fig:prog_eg}. $t_0$ denotes the time upon which the explosion shock breaks out of the stellar surface. }
    \label{fig:lc_eg}
\end{figure}

\begin{figure}
    \centering
    \includegraphics[width=0.48\textwidth]{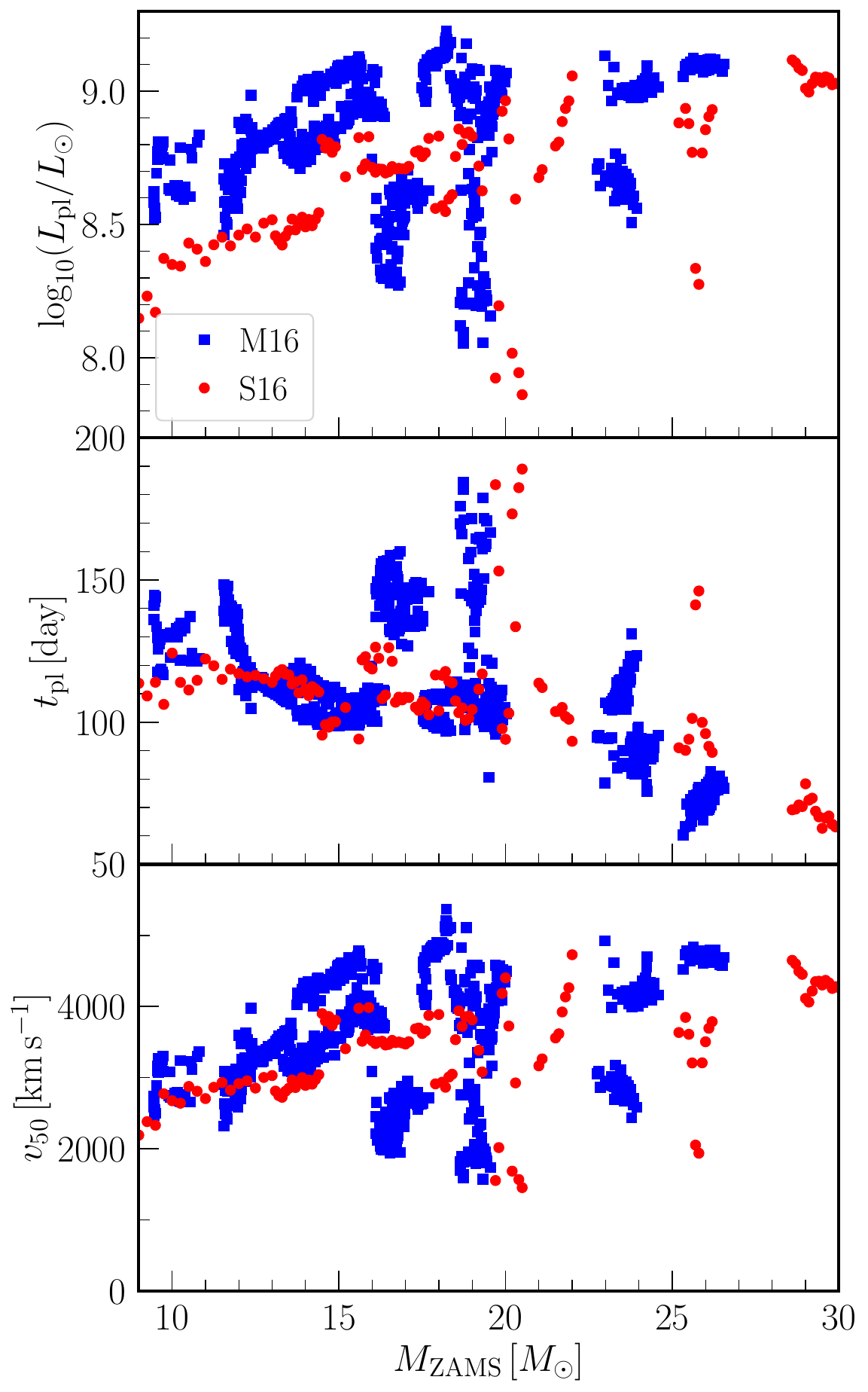}
    \caption{Comparison of light curve parameters as a function of ZAMS mass between M16 (\textsl{blue squares}) and S16 (\textsl{red dots}) models.  From top to bottom, the panels show the plateau luminosity, plateau duration and photospheric velocity at 50 days after the shock breakout, respectively.}
    \label{fig:lc_zams}
\end{figure}

As representative examples, we plot in Fig.~\ref{fig:lc_eg} the bolometric LCs of SNe IIP from the pre-SN models shown in Fig.~\ref{fig:prog_eg}, with their respective $L_\mathrm{pl}$ and $t_\mathrm{pl}$ given in Table~\ref{tab:prog_eg}.  It is clear at a first glance that the M16 models are brighter than S16 models during the plateau phase for the same $M_{\rm ZAMS}$, despite the similar explosion properties (also listed in Table~\ref{tab:prog_eg}).  This feature is further exemplified in Fig.~\ref{fig:lc_zams}, which compares $L_{\rm pl}$, $t_{\rm pl}$ and $v_{50}$ as a function of $M_{\rm ZAMS}$ between all M16 and S16 models that successfully explode.  Whereas $t_{\rm pl}$ is quite similar for the two sets of models, $L_{\rm pl}$ of M16 models is in general larger by a factor of $\mathord\sim2$ than that of S16 models.  This difference cannot be accounted for even by appealing to large uncertainties in the explosion energy.  Similar values of $L_{\rm pl}$ as in S16 can only be realized for M16 models by artificially dividing $E_{\rm exp}$ by three, which is unrealistic and would affect $t_{\rm pl}$ considerably.  Indeed, the difference in $L_{\rm pl}$ reflects the systematically different envelope structure between M16 and S16 progenitors (see the density profiles in Fig.~\ref{fig:prog_eg} and the pre-SN masses and radii in Fig.~\ref{fig:prog_props}).  The slightly larger $v_{50}$ for M16 models reflects the smaller $M_{\rm prog}$ of the M16 progenitors which leads to a smaller $M_{\rm ejecta}$ (cf. panel (a) in Fig.~\ref{fig:prog_props}).  As we shall see in \S\ref{sec:result}, the comparison with observations suggests a preference for the S16 models as realistic progenitors as they match the observed plateau luminosities better.

\begin{figure*}
    \centering
    \includegraphics[width=0.96\textwidth]{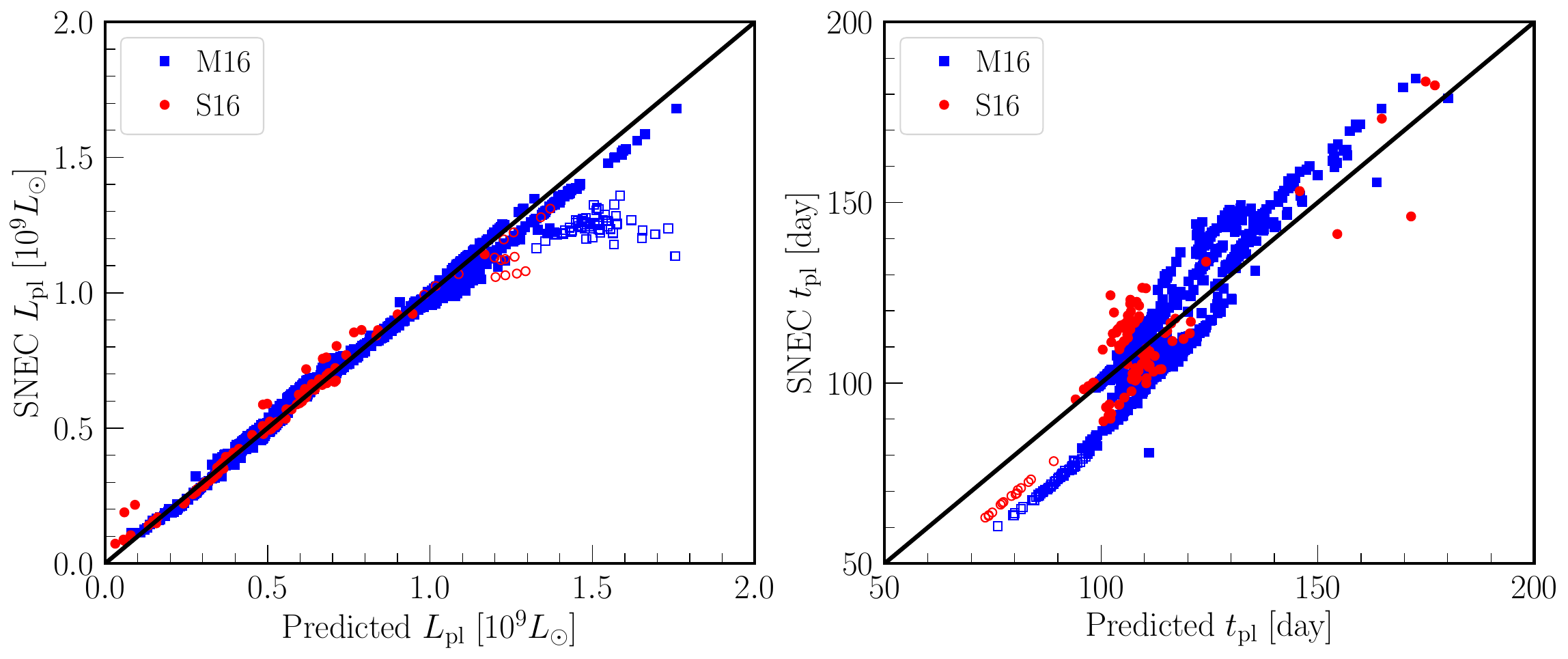}
    \caption{Comparison of light curve parameters, i.e., plateau luminosity ($L_\mathrm{pl}$, \textsl{left panel}) and duration ($t_\mathrm{pl}$, \textsl{right panel}), between SNEC simulations (\textsl{ordinate}) and the analytic scaling relations in Eqs.~(\ref{eq:lpl}) and (\ref{eq:tpl}, \textsl{abscissa}) for M16 and S16 progenitors.  \textsl{Open symbols} indicate the models with $t_\mathrm{pl}\le80\,\rm days$, which leads to discrepancy of $L_\mathrm{pl}$ between \texttt{SNEC} results and the scaling relation for M16 models.  The \textsl{black lines} in both panels mark the diagonal.}
    \label{fig:scaling}
\end{figure*}

Lastly, we compare our results to analytic scaling relations often used by observers to infer the properties of progenitor and explosion from LC parameters, both to guide the interpretation of our results and to check the validity of the analytic relations.  For $L_{\rm pl}$, we use the relation derived in \cite{1993ApJ...414..712P}
\begin{equation} \label{eq:lpl}
    L_{\rm pl} = L_0 E_{51}^{5/6} M_{10}^{-1/2} R_{\rm 0,500}^{2/3},
\end{equation}
where $E_{51}$ is the explosion energy in units of $10^{51}$\,erg, $M_{10}$ is the mass of the hydrogen envelope (the progenitor mass minus the helium core mass) in units of 10\,$M_\odot$, and $R_{\rm 0,500}$ is the pre-SN stellar radius $R_\mathrm{prog}$ in units of $500\,R_\odot$.  Our preferred values of $L_0$ are $1.69\times10^{42}\,$erg\,s$^{-1}$ and $1.51\times10^{42}\,$erg\,s$^{-1}$ for M16 and S16 models, respectively.  The left panel of Fig.~\ref{fig:scaling} shows that Eq.~(\ref{eq:lpl}) predicts $L_{\rm pl}$ well overall, with a relative error $\lesssim10\%$ for most models. The discrepancy for models with a large $L_{\rm pl}$ with a relative error up to $40\%$ is due to their short plateau for which $L_{\rm bol}$ at 50 days may not well represent $L_{\rm pl}$. 

The scaling relation for the plateau duration from \cite{1993ApJ...414..712P} assumes no energy input from radioactive decay of nickel and cobalt and reads
\begin{equation}
    t_{\rm pl,0} = t_0 \, E_{51}^{-1/6} M_{10}^{1/2} R_{\rm 0,500}^{1/6}.
\end{equation}
Following \cite{2016ApJ...821...38S}, we use a modified relation for $t_{\rm pl}$ that takes into account that energy input from radioactive decay can prolong the plateau, 
\begin{equation} \label{eq:tpl}
\begin{aligned}
    t_{\rm pl} &= t_{\rm pl,0}\times f_{\rm rad}^{1/6}, \\
    f_{\rm rad} &= 1+C_f M_{\rm Ni}E_{51}^{-1/2}M_{10}^{-1/2}R_{\rm 0,500}^{-1}, 
\end{aligned}
\end{equation}
where we set the constant $C_f=21$ as suggested in \cite{2016ApJ...821...38S}.
Comparing the LCs from SNEC to Eq.~(\ref{eq:tpl}) is more appropriate, as SNEC includes the energy release from radioactive decay.  The fitted $t_0$ are $93.0\,\mathrm{d}$ and $89.7\,\mathrm{d}$ for M16 and S16 models, respectively.  The right panel of Fig.~\ref{fig:scaling} shows that Eq.~(\ref{eq:tpl}) predicts $t_{\rm pl}$ well at $t_{\rm pl}\gtrsim100$\,days, with a relative error $\lesssim15\%$.  For $t_{\rm pl}\lesssim100$\,days, the relative error can be up to $\sim25\%$.

\section{Comparison to observations} \label{sec:result}
\subsection{Global statistics}
Our large ensemble of stellar models allows for a statistical comparison to observational data.  As a first step towards such a quantitative comparison, we choose the volume-limited set of well-observed nearby SNe IIP from PP15, who provide $L_{\rm pl}$, $t_{\rm pl}$, and $M_{\rm Ni}$, using their own LC fitting method consistently across the photometric data of the entire sample instead of just collecting LC parameters from the literature.  Following \cite{2015ApJ...806..225P}, we use a subset from the PP15 sample including 17 SNe IIP with well-determined photometry\footnote{\citet{2015ApJ...806..225P} include SN2013am in their analysis, but no quantitative results were given for this particular SN. Also, we exclude SN1980K, which is a Type IIL.}. 

\begin{deluxetable*}{cccccccccc}
\tablecaption{Global statistical parameters of light curves from observations and theoretical models.} \label{tab:lc_mean}
\tablehead{\colhead{Data set} & \multicolumn{3}{c}{{$\log_{10}(L_{\rm pl}/L_\odot$)}} &
\multicolumn{3}{c}{$t_{\rm pl}$~(day)} & \multicolumn{3}{c}{$\log_{10}(M_{\rm Ni}/M_\odot)$} \\
{} & \colhead{Mean} & \colhead{$\sigma$} & \colhead{$p$-value} & \colhead{Mean} & \colhead{$\sigma$} & \colhead{$p$-value} & \colhead{Mean} & \colhead{$\sigma$} & \colhead{$p$-value}}
\startdata
PP15 & 8.39 & 0.39 & --- & 119 & 13 & --- & -1.52 & 0.48 & --- \\
M16  & 8.76 & 0.17 & $4\times10^{-8}$ & 123 & 16 & 0.04 & -1.37 & 0.19 & 0.005 \\
S16  & 8.49 & 0.23 & 0.21 & 113 & 13 & $2\times10^{-4}$ & -1.35 & 0.22 & 0.03\\
\enddata
\end{deluxetable*}

\begin{figure*}
    \centering
    \includegraphics[width=0.92\textwidth]{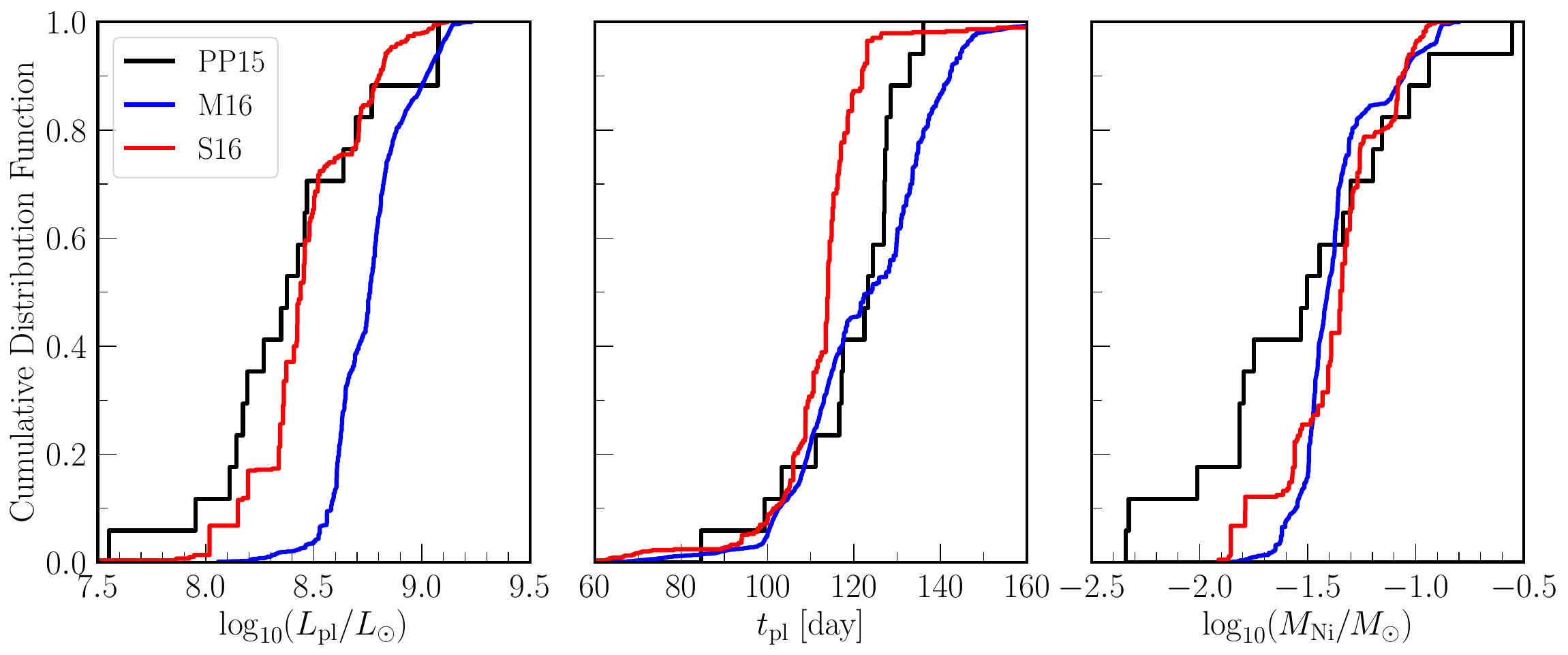}
    \caption{Comparison of the cumulative distribution function of the light curve parameters between the theoretical model sets (M16 and S16) and the observational data set (PP15). }
    \label{fig:CDF}
\end{figure*}

Here, we compare global statistical parameters in theoretical models to observations.  For theoretical model sets, we calculate the weighted means of the LC parameters, defined as
\begin{equation} \label{eq:mean}
    \langle a\rangle = \frac{\sum_i  a_i w(M_i)\Delta M_i}{\sum_i w(M_i)\Delta M_i}\;.
\end{equation}
Here, $a$ stands for any of the variables $\log_{10}(L_\mathrm{pl}/L_\odot)$, $t_\mathrm{pl}$, or $\log_{10}(M_\mathrm{Ni}/M_\odot)$.  The Salpeter initial mass function (IMF, \citealp{1955ApJ...121..161S}) is used as the weighting function, i.e., $w(M_i) \propto M_i^{-2.35}$, and $\Delta M_i$ is the resolution of the ZAMS mass grid around $M_i$. 
We set the minimum and maximum $M_i$ to $9\,M_\odot$ and $30\,M_\odot$, respectively.  For the observational data, we give each SN the same weight as appropriate for a volume-limited sample.  The standard deviation $\sigma$ of the LC parameters is evaluated as
\begin{equation} \label{eq:sd}
    \sigma = \sqrt{\frac{\sum_i (a_i-\langle a\rangle)^2 w(M_i)\Delta M_i }{\sum_i w(M_i)\Delta M_i }}.
\end{equation}
The M16 set has a deficit of models with $M_\mathrm{ZAMS}$ from $9\,M_\odot$ to $12\,M_\odot$ (see gaps in Fig.~\ref{fig:prog_props}).  The pre-SN evolutionary simulation of stars near the low-mass end is difficult and beset with uncertainties due to the increasing influence of degeneracy in the core \citep{2015ApJ...810...34W}, and awaits for further improvement.  To accommodate the deficit of low-mass models, we assign the weight in a $1\,M_\odot$ bin to the existing models
\begin{equation} \label{eq:m16_low}
    w(M) = w_0(M) \frac{\int_{M_0}^{M_0+1M_\odot}w_0(M'){d}M'}{\sum\limits_{M_i \in [M_0, M_0+1M_\odot]} w_0(M_i)\Delta M_i}, 
\end{equation}
where $w_0(M)$ is the original weight from the IMF and $M_0=9,\,10\,,11\,M_\odot$.

Table~\ref{tab:lc_mean} summarizes the global statistical parameters for the LCs from the two theoretical model sets and the PP15 sample.  This is supplemented by the cumulative distribution functions (CDF) of the LC parameters as shown in Fig.~\ref{fig:CDF}.  Due to generally smaller progenitor radii and slightly higher envelope masses, the S16 models generally have a lower $L_{\rm pl}$ that better agrees with the PP15 sample. However, the CDF of theoretical $L_{\rm pl}$ shows a deficit of models with low luminosity $L_{\rm pl}\le10^8\,L_\odot$.  M16 models give a longer mean plateau duration of $\sim123$\, days because low-mass models ($M_\mathrm{ZAMS}\le12\,M_\odot$) have $t_{\rm pl}\ge120\,$days (Fig.~\ref{fig:lc_zams}).  The comparison of the CDF of $t_{\rm pl}$ shows both theoretical models struggle to reproduce all the observational constraints. However, this discrepancy may partly be due to the different definition of $t_{\rm pl}$ between this work and PP15.  For $M_{\rm Ni}$, M16 and S16 models give very similar mean values and CDFs.  This is expected as $M_{\rm Ni}$ mainly depends on the core structure and the explosion model, which are similar in both model sets (Fig.~\ref{fig:exp}).  Comparing to the PP15 data, our theoretical models have a slightly larger mean $M_{\rm Ni}$, and, based on the CDF, this is likely due to a lack of models with very small nickel masses $M_\mathrm{Ni}<0.01\,M_\odot$.  The scarcity of models with low nickel mass and (to a lesser extent for the S16 models) low luminosity, might be due to the absence of electron-capture supernovae in the model sets \citep{2021MNRAS.503..797K,2022MNRAS.513.1317Z}\footnote{Note, however, that adding electron-capture supernovae may only help to add explosions with low nickel mass, but not with low luminosity \citep{moriya_14}.}; the mass range for electron-capture supernovae remains quite uncertain \citep{2017PASA...34...56D,2008ApJ...675..614P}.
Another cause could be uncertainties for models with near-critical explodability (the gray shaded regions in Fig.~\ref{fig:exp_S16}). It is possible that some of these models might result in low-energy explosions that produce little nickel and may experience fallback (which could remove nickel as well).
 
To further assess discrepancies between the observed and predicted distribution of explosion properties, we perform individual Kolmogorov-Smirnov (K-S) tests for each LC parameter to estimate the goodness of fit of our theoretical models to the PP15 sample.  For each theoretical model set, we generate a large random sample of SN IIP models following the IMF.  For the M16 set, we assign the weight for $M_{\rm ZAMS}$ below $12\,M_\odot$ to the existing models according to Eq.~(\ref{eq:m16_low}).  We choose a sample size of $10^5$ so that the random sample well reproduces the theoretical CDFs.  The large sample size ensures that the random generation process does not affect the resultant $p$-values of K-S tests, which are listed in Table~\ref{tab:lc_mean}.  The K-S test for $L_{\rm pl}$ suggests an obvious preference of S16 models over M16 models, agreeing with our assessment of the mean $L_{\rm pl}$.  The K-S test for $t_{\rm pl}$ favors the M16 models, but the fit is far from perfect with indications of possibly significant differences to the observed distribution ($p$-value of $0.04$).  Note that the test statistic is subject to uncertainties in obtaining $t_{\rm pl}$ for both models and observations.  As expected from the lack of models with low $^{56}$Ni yields and smaller mean $M_{\rm Ni}$ in our models, the K-S tests for $M_{\rm Ni}$ show both model sets struggle to fit the PP15 sample. 

\subsection{Correlations between explosion properties}
\begin{figure}
	\centering
	\includegraphics[width=0.47\textwidth]{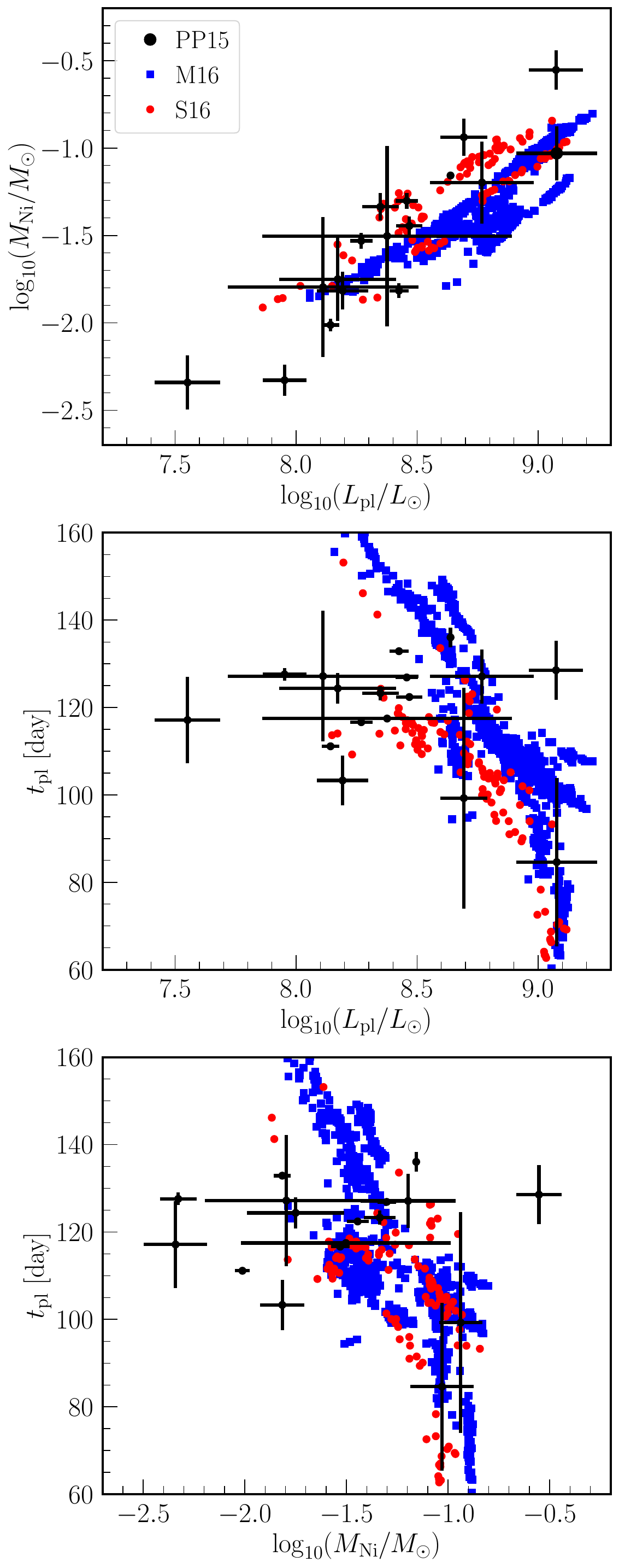}
	\caption{Correlations between light curve parameters for theoretical predictions of the M16 (\textsl{blue squares}) and S16 (\textsl{red dots}) models and the PP15 data set (\textsl{black dots}).}
	\label{fig:corr}
\end{figure}

\begin{figure}
    \centering
    \includegraphics[width=0.48\textwidth]{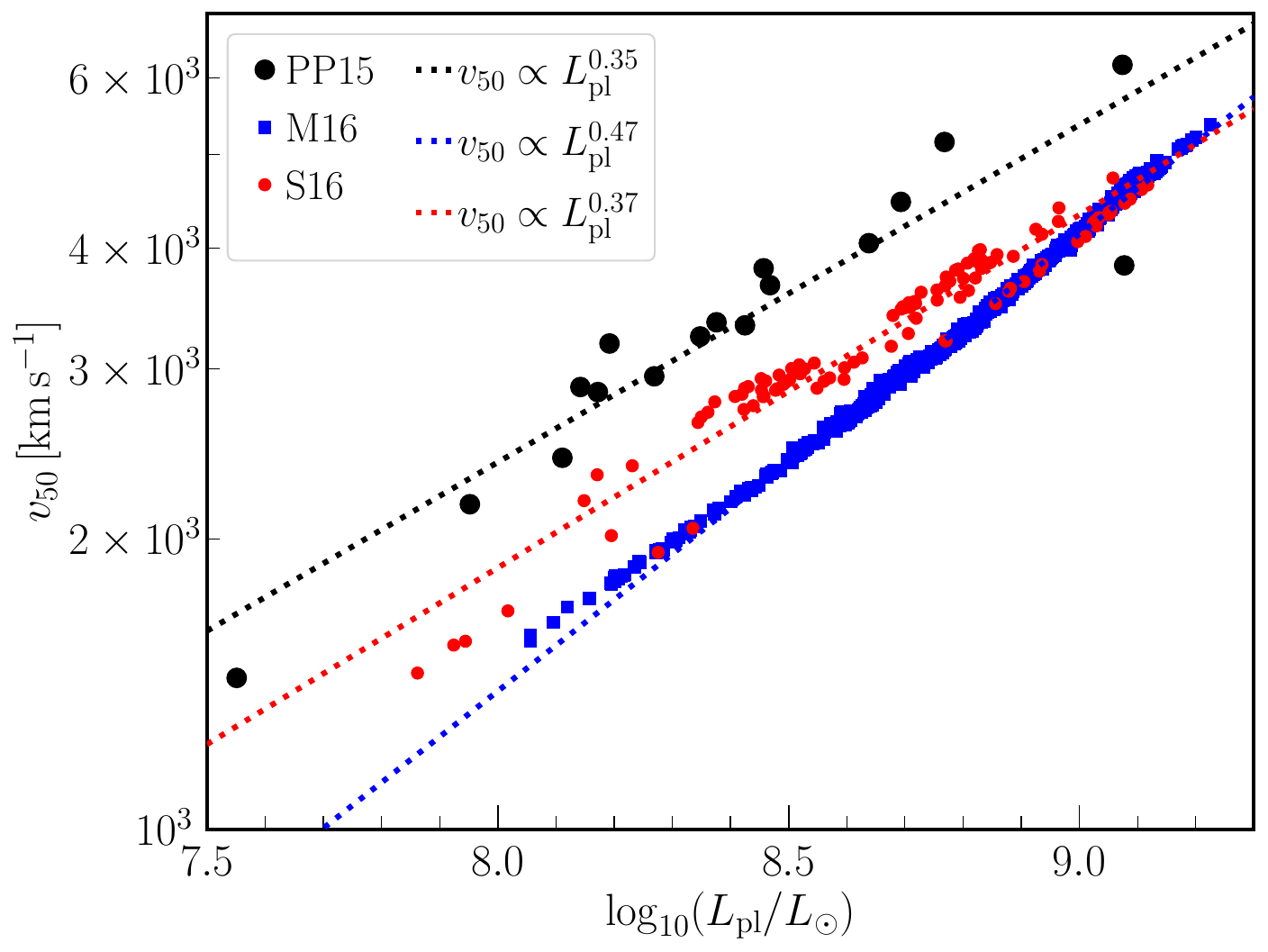}
    \caption{Correlation between the plateau luminosity and photospheric velocity at 50 days after the shock breakout for theoretical predictions of the M16 (\textsl{blue squares}) and S16 (\textsl{red dots}) models and the PP15 data set (\textsl{black dots}). The dashed curves represent the best fit with a power-law relation for the corresponding data points.}
    \label{fig:v50_lpl}
\end{figure}

\begin{deluxetable*}{cccc}
\tablecaption{Correlation matrix elements of the LC parameters in the observational sample and our model sets.} \label{tab:lc_covar}
\tablehead{\colhead{Data set} & \colhead{$\rho(\log_{10}(L_{\rm pl}/L_\odot),\log_{10}(M_\mathrm{Ni}/M_\odot))$} & \colhead{$\rho(\log_{10}(L_{\rm pl}/L_\odot),t_{\rm pl})$} & \colhead{$\rho(\log_{10}(M_\mathrm{Ni}/M_\odot),t_{\rm pl})$}} 
\startdata
PP15 & 0.92 & -0.18 & -0.12  \\
M16  & 0.85 & -0.91 & -0.71 \\
S16  & 0.83 & -0.61 & -0.41 \\
\enddata
\end{deluxetable*}

Correlations have been found between LC parameters in observations and inferred explosion properties \citep[e.g., $L_{\rm pl}$ and $M_{\rm Ni}$, see][]{hamuy_03,poznanski_12,chugai_14,2015ApJ...806..225P,mueller_t_17}.  Correlations can also allow to put constraints on the theoretical progenitor and explosion models. 
Figure~\ref{fig:corr} shows three pairs of LC parameters from the PP15 sample and the two model sets in this work.  Visually, one can see that both M16 and S16 model sets possess correlations between all three pairs of parameters, whereas PP15 only exhibits a clear correlation between $L_{\rm pl}$ and $M_{\rm Ni}$.  Comparison of the theoretically predicted two-dimensional distribution of $L_{\rm pl}$ and $M_{\rm Ni}$ to that in the PP15 sample also suggests a preference for S16 models due to their smaller $L_{\rm pl}$, agreeing with the conclusion drawn from the global statistical parameter. 

To quantify the strength of the predicted and observed correlations, we calculate the weighted correlation matrix elements as
\begin{equation*}
    \rho(a,b) = \frac{\sum_i (a_i-\langle a\rangle)(b_i-\langle b\rangle)w(M_i)\Delta M_i }{\sigma_a \sigma_b\sum w(M_i)\Delta M_i  },
\end{equation*}
for any pair of parameters $a$ and $b$, and $i$ runs over all data/bins. 
Here we take $\log_{10}(L_\mathrm{pl}/L_\odot)$, $t_\mathrm{pl}$, and $\log_{10}(M_\mathrm{Ni}/M_\odot)$ as the LC parameters.  Similar to our analysis in the previous section, we use the Salpeter IMF as the weight $w$ for theoretical models and assign the same weight for each SN in the PP15 sample.  The three non-trivial correlation matrix elements for PP15, M16, and S16 are given in Table~\ref{tab:lc_covar}.  The correlation between $L_{\rm pl}$ and $M_{\rm Ni}$ are similar between either of our model sets and the PP15 sample, while the pronounced correlation between $L_{\rm pl}$ and $t_{\rm pl}$ found in both sets is clearly absent in the PP15 sample.  This discrepancy indicates the need for further investigation with a larger SN IIP sample.  Although the presence or absence of a correlation may be somewhat altered by a more consistent determination of $t_{\rm pl}$ in models and observations, the discrepancy may indicate missing physics in the explosion models or the progenitor structure.  Specifically, the effect of adding Type~IIP progenitors that have undergone binary interactions \citep{podsiadlowski_92,zapartas_21} needs to be investigated.  Although it is plausible that binary interactions could destroy the predicted correlation between $L_{\rm pl}$ and $t_{\rm pl}$ (which may be spurious), it is not clear how binary effects could reduce the overly large spread in $t_{\rm pl}$; in fact they might even exacerbate this problem.

\citet{2002ApJ...566L..63H} presented an interesting correlation between expansion velocities of SNe IIP ejecta during plateau phase and the plateau luminosity, which can make SNe IIP standardized candles. In Fig.~\ref{fig:v50_lpl} we present a similar $v_{50}-L_{\rm pl}$ diagram for our theoretical model sets M16 and S16 and the PP15 observation sample. Note that $v_{50}$ for the PP15 sample were estimated by Eq.~(3) of \citet{2015ApJ...799..215P} which was used to fit available FeII velocities, while for theoretical models, $v_{50}$ stands for the velocity of the photosphere defined by $\tau=2/3$. Further spectroscopic modelling is needed to better understand the discrepancy between models and observations \citep{2011MNRAS.410.1739D}. 
It is possible that some of the discrepancy between models and observations is due to the use of 
a proxy for the FeII velocity in the models instead of determining it by detailed spectral radiative transfer.
If the values of $v_{50}$ are taken at face value, we observe a systematic difference for the $v_{50}-L_{\rm pl}$ relation between our two model sets, which is primarily due to the difference in their $L_{\rm pl}$. This again exemplifies the power of using correlations between SN observables to constrain theoretical models.

\section{Conclusions} \label{sec:conclu}

In this paper, we presented $\sim1100$ light curves of SNe Type IIP generated by \texttt{SNEC} from two sets of single-star solar-metallicity progenitor models in M16 \citep{2016MNRAS.460..742M} and S16 \citep{2016ApJ...821...38S}, with very high resolution in ZAMS mass grid as fine as $0.01\,M_\odot$ in the former set.  We assume that SNe IIP are driven by neutrinos and calculate the key explosion parameters $E_{\rm exp}$, $M_{\rm NS,by}$ and $M_{\rm Ni}$ using a semi-analytical approach derived in \cite{2016MNRAS.460..742M}.

The explosion parameters agree well globally between the M16 and S16 model sets and between the semi-analytic model and alternative phenomenological explosion models from previous studies of exploding S16 models \citep{2016ApJ...821...38S,2022ApJ...934...67B}.  In particular, the agreement between the prediction of the semi-analytic model and the 1D simulations of \citet{2016ApJ...821...38S} for the same progenitor set is striking.  The plateaus of SNe Type IIP are systematically fainter by a factor of $\mathord\sim2$ in bolometric luminosity for the S16 set due to denser hydrogen envelopes of S16 progenitors.  The more extended envelope structure of the M16 models lead to brighter plateaus and is likely artificial because of simplification of the surface boundary condition in the stellar evolution calculations.  This reinforces previous findings on the sensitivity of Type~IIP explosions to the envelope structure \citep{dessart_13} and implies that difference in theoretical light curves may rather reflect assumptions about stellar structure and evolution, in particular those that affect the structure of the convective RSG envelope, than the modeling of the explosion engine.  As already pointed out by  \citet{dessart_19}, this may cause problems in inferring progenitor properties from observables, e.g., inferring the ZAMS mass from the plateau luminosity \citep{2022arXiv221105789B}.  It is important to highlight that even among available stellar evolution models computed with the same code, there may be subtle different in the treatment of the convective envelope and outer boundary due to code improvements and model parameter choices that may have significant repercussions for supernova light curve modeling.  To fully exploit the diagnostic potential of SNe Type~IIP light curves, more theoretical and observational work on RSG envelopes and environments is critical.

We compare the parameters of the predicted light curves to the volume-limited PP15 sample of well-observed SNe IIP \citep{2015ApJ...799..215P}.  We construct a mock supernova population from the two progenitor sets by weighting the models with the Salpeter IMF.  Based on the mean value of the plateau luminosity $L_{\rm pl}$ and a K-S test, the S16 models fit the observed brightness distribution in the PP15 SNe IIP sample better.  We find a similar correlation between $L_{\rm pl}$ and $M_{\rm Ni}$ in both model sets and in the PP15 sample. 
However, we find tensions with the observational data for both model sets, which may either indicate an incomplete understanding of the progenitor-explosion connection or of the pre-supernova progenitor structure.  Both progenitor sets lack models with explosions that produce the very small nickel masses $M_{\rm Ni}<0.01\,M_\odot$ that are observed in some IIP explosions.  This discrepancy may be related to the uncertainties in models at the low-mass end \citep{2015ApJ...810...34W} or to models close to black-hole formation.  The comparison of the plateau duration $t_{\rm pl}$ remains beset with ambiguities in the definition of the plateau length, but we tentatively find a significantly larger spread in plateau duration in the models compared to the observed sample.  Furthermore, the models predict an anti-correlation between $L_{\rm pl}$ and $t_{\rm pl}$, which is not found in the PP15 sample.  Indications of an anti-correlation have, however, been found in other samples \citep{faran_14}, and future studies need to assess whether bigger volume-limited samples confirm the tension between theory and observations.

These results provide an interesting lead for further comparisons of the theoretical models and observational data for SNe IIP LCs.  In particular, the predicted correlation between plateau luminosity and plateau duration would present a challenge to current stellar evolution models of massive stars, if the tension to observations can be corroborated using larger volume-limited transient samples.  Future studies should explore variations of single-star and binary evolution models and pit them against bigger volume-limited transient samples from recent and upcoming surveys.  By obviating the need for time-critical 1D (let alone 3D) supernova simulations, our semi-analytic model may be useful for conducting such large-scale comparisons more efficiently with little loss of accuracy, given the remarkable agreement with the explosion properties obtained by \citet{2016ApJ...821...38S}.

\begin{acknowledgments}
We thank Evan O'Connor for useful discussion. SZ is supported by the China Postdoctoral Science Foundation (2022M712082). The simulations were run on the Siyuan-1 cluster supported by the Center for High Performance Computing at Shanghai Jiao Tong University.  BM was supported by ARC Future Fellowship FT160100035.  BM and AH are supported by the Australian Research Council (ARC) Centre of Excellence (CoE) for Gravitational Wave Discovery (OzGrave) project number CE170100004.  AH is supported by the ARC CoE for All Sky Astrophysics in 3 Dimensions (ASTRO 3D) project number CE170100013.  We acknowledge computer time allocations from Astronomy Australia Limited's ASTAC scheme and the National Computational Merit Allocation Scheme (NCMAS), and from and Australasian Leadership Computing Grant.
\end{acknowledgments}

\software{\texttt{SNEC} \citep{2015ApJ...814...63M}, \texttt{NumPy} \citep{numpy}, \texttt{Matplotlib} \citep{matplotlib}}, \texttt{SciPy} \citep{2020SciPy-NMeth}

\bibliography{the_bib}{}
\bibliographystyle{aasjournal}



\end{document}